\documentclass[aps,floats,eqsecnum,draft,showpacs,superscriptaddress]{revtex4}
\usepackage{psfig}
\usepackage{amsmath}

\begin{document}

\def\d{\partial}
\def\w{\omega}
\def\go1{g_0^{(1)}}
\def\gon{g_0^{(N)}}
\def\goj{g_0^{(j)}}
\def\g{\gamma}
\def\b{\beta}
\def\a{\alpha}
\def\l{\Lambda}
\def\lan{\lambda_n}
\def\bk{\frac{\b^2}{k}}
\def\NI{\noindent}
\def\qh{{\hat q}}
\def\Qh{{\hat Q}}
\def\gh{{\hat g}}
\def\z{{\overline{z}}}
\def\k{{\overline{k}}}
\def\half{{1\over2}}
\def\kR{k_{\rm R}}
\def\kI{k_{\rm I}}
\def\C{{\rm\bf C}}
\def\O{{\cal O}}
\def\phi{\varphi}
\def\kj{\kappa_j}
\def\mj{\mu_j}
\def\lj{\lambda_j}
\def\und#1{$\underline{\smash{\hbox{#1}}}$}
\def\e{e^{ikx-\w(k)t}}
\def\edot{e^{ik\cdot x-\w(k)t}}
\def\Im{{\rm Im}\,}
\def\Re{{\rm Re}\,}
\def\t{\tau}
\def\cb{C^{\beta}}

\title{A Generalization of both the Method of Images and of the Classical
Integral Transforms}

\author{Athanassios S. Fokas\footnote{Permanent address:
Department of Applied Mathematics and Theoretical Physics, Cambridge University, Cambridge CB3 0WA, UK. }}\email{T.Fokas@damtp.cam.ac.uk}
\affiliation{Institute for Nonlinear Studies, Clarkson University, Potsdam, NY
13699-5805}
\author{Daniel ben-Avraham}\email{benavraham@clarkson.edu}
\affiliation{Physics Department, Clarkson University, Potsdam, NY 13699-5820}

\begin{abstract}
A new method for the solution of initial-boundary value problems
for evolution PDEs recently introduced by Fokas is generalised to
multidimensions. Also the relation of this method with the method of
images and with the classical integral  transforms is discussed. The
new method is easy to implement, yet it is applicable to problems
for which the classical approaches apparently fail. As illustrative
examples, initial-boundary value problems for the diffusion-convection
equation in one and higher dimensions, as well as for the linearised 
Korteweg-de Vries equation with the space variables on the half-line
are solved. The suitability of the new method for the analysis of the
the long-time asymptotics is ellucidated.
\end{abstract}


\maketitle

This paper is dedicated to J.B.Keller on the occasion of his 80th
birthday.

\section{introduction}

The goal of this paper is to extend to multidimensions a new method
recently introduced by Fokas~\cite{IMA,JMP}, as well as to discuss
the relation of this method with the classical methods of
images and of integral transforms.  It will be shown that the new method
provides an appropriate generalization of the classical approaches.

For simplicity, we will limit our discussion to evolution PDEs on the
half line. Evolution PDEs on a finite domain are discussed in
\cite{fokpel,pell}. The new method can also be applied to elliptic
PDEs \cite{fokas1}, such as the Laplace~\cite{F-Kapaev}, 
Helmholtz~\cite{F-DbA}, and biharmonic equations~\cite{F-Crowdy}. The
extension of this method to nonlinear integrable evolution PDEs is
discussed in \cite{JMP,fokas2}.

In order to help the reader become
familiar with the new method, rather than discussing general
initial-boundary value (IBV) problems, we will
concentrate on the following three concrete, physically significant
problems:
\begin{enumerate}
\item
The diffusion-convection equation,
\begin{subequations}
\label{1.1} 
\begin{eqnarray}
&&q_t=q_{xx}+\a q_x,\qquad 0<x<\infty,\quad t>0,\\
&&q(x,0)=q_0(x),\qquad0<x<\infty,\\
&&q(0,t)=g_0(t),\qquad\>\> t>0,
\end{eqnarray}
\end{subequations}
where $\a$ is a real constant. 
\item
 The linearized Korteweg - de Vries equation (with dominant surface tension),
\begin{subequations}
\label{1.2}
\begin{eqnarray}
&&q_t+q_x-q_{xxx}=0,\qquad0<x<\infty,\quad t>0,\\
&&q(x,0)=q_0(x),\qquad\>\>\>\>\>\>\,0<x<\infty,\\
&&q(0,t)=g_0(t),\quad q_x(0,t)=g_1(t),\qquad t>0.
\end{eqnarray}
\end{subequations}
\item
The multidimensional diffusion-convection equation,
\begin{subequations}
\label{1.3}
\begin{eqnarray}
&&q_t=\sum_{j=1}^N(\d_{x_j}^2+\a_j\d_{x_j})q,\qquad 0<x_j<\infty,\quad
j=1,2,\dots,N,\quad t>0,\\
&&q(x_1,x_2,\dots,x_N,0)=q_0(x_1,x_2,\dots,x_N),\qquad0<x_j<\infty,\quad
j=1,2,\dots,N,\\
&&q(0,x_2,\dots,x_N,t)=\go1(x_2,\dots,x_N,t),\qquad\qquad\qquad\>\,
0<x_j<\infty,\quad j=2,\dots,N,\qquad\,\,\,\, t>0,\nonumber\\
&&\qquad\vdots\nonumber\\
&&q(x_1,x_2,\dots,x_{N-1},0,t)=\gon(x_1,x_2,\dots,x_{N-1},t),
\qquad0<x_j<\infty,\quad j=1,\dots,N-1,\quad t>0,
\end{eqnarray}
\end{subequations}
where $\a_j$, $j=1,\dots,N$ are real constants. The functions $q_0$, $g_0$,
$g_1$, $\{\goj\}_1^N$ have appropriate smoothness and they also decay
as $x$ and $x_j$ tend to $\infty$.
\end{enumerate}

The discussion of the physical significance of the above IBV problems, as
well as the derivation of their solution is presented in sections~II--IV.

\bigskip\noindent
{\bf The proper transform in $(x,t)$}
\medskip

The proper transform of a given IBV problem is specified by the PDE, the
domain, and the boundary conditions.  For {\em simple} IBV problems there
exists an algorithmic procedure for deriving the associated transform
(see, for example, \cite{fried,stak}).  This procedure is based
on separating variables and on analyzing {\em
one} of the resulting eigenvalue equations.  Thus, for simple IBV problems
in $(x,t)$ there exists a proper $x$-transform and a proper
$t$-transform.  Sometimes these transforms can be found by inspection. For
example, for the IBV~(\ref{1.1}) with $\a=0$ the proper $x$-transform is
the sine transform, and if $0<t<\infty$ the proper $t$-transform is the
Laplace transform.

For a general evolution equation in $\{0<x<\infty,t>0\}$, the
$x$-transform is more convenient than the $t$-transform.  For example,
looking for a solution of the form $\exp[ikx-\w(k)t]$ in Eqs.~(\ref{1.1}a)
and (\ref{1.2}a), we find that $\w(k)$ is given {\em explicitly} by
\begin{equation}
\w(k)=k^2-i\a k,\qquad \w(k)=i(k+k^3).
\end{equation}
On the other hand, looking for solutions of the form
$\exp[-st+\lambda(s)x]$ we find that $\lambda(s)$ is given only {\em
implicitly}, by
\begin{equation}
\label{1.5}
\lambda^2+\a\lambda+s=0,\qquad\lambda^3-\lambda+s=0.
\end{equation}
The advantage of the $x$-transform for an evolution equation becomes clear
when the domain is the infinite line $-\infty<x<\infty$, in which case the
$x$-Fourier transform yields an elegant representation for the initial
value problem of an {\em arbitrary} evolution equation.  Furthermore, {\em
if an} x-{\em transform exists}, it provides a
convenient method for the solution of problems defined on the half line
$0<x<\infty$. However, in general an $x$-transform does {\em not} exist;
this is, for example, the situation for the IBV problem~(\ref{1.2}).  In
this case, until recently one had no choice but to attempt to use the
$t$-Laplace transform.  For example, Eqs.~(\ref{1.2}) yield
\begin{subequations}
\label{1.6}
\begin{eqnarray}
&&-{\tilde q}_{xxx}+{\tilde q}_x+s{\tilde q}=q_0(x),\\
&&{\tilde q}(0,s)={\tilde g}_0(s), \qquad {\tilde q}_x(0,s)={\tilde
g}_1(s),
\end{eqnarray}
\end{subequations}
where ${\tilde q}(x,s)$, ${\tilde g}_0(s)$, ${\tilde g}_1(s)$, denote the
Laplace transforms of $q(x,t)$, $g_0(t)$, $g_1(t)$, respectively.  However,
this approach is rather problematic: (a)~The problem is posed for finite
$t$, say $0<t<T$, while the Laplace transform involves $0<t<\infty$. 
Thus, the correct transform is
\[
{\tilde q}(x,s)=\int_0^Tdt\,e^{-st}q(x,t).
\]
This yields the additional term $-q(x,T)\exp(-st)$ in the r.h.s. of
Eq.~(\ref{1.6}a); some authors use ``causality" arguments to justify
replacing $T$ by $\infty$. (b)~If $T=\infty$ and if $g_0(t)$, $g_1(t)$ decay
for large $t$, then the application of the Laplace transform can be
justified.  However, since the homogeneous version of (\ref{1.6}a)
involves $\exp(-\lambda(s)x)$, where $\lambda$ solves the {\em cubic}
equation~(\ref{1.5}b), the investigation of Eqs.~(\ref{1.6}) is cumbersome.

\bigskip\noindent
{\bf The method of images}
\medskip

A broad class of IBV problems is often approached by the method of images.  Suppose that~\cite{remark}
\begin{subequations}
\label{d1}
\begin{eqnarray}
&&q_t+\w(i\d_x)q=0,\qquad\qquad 0<x<\infty,\quad t>0,\\
&&q(x,0)=q_0(x),\qquad\qquad\quad\! 0<x<\infty,\\
&&q(0,t)=g_0(t),\qquad\qquad\>\>\>\>\,\, t>0,
\end{eqnarray}
\end{subequations}
where $\w(\cdot)$ is an {\em even} polynomial of its argument.  The method of images then yields a solution as follows.  Let $G(x,x',t)$ be the Greens function that satisfies
\begin{eqnarray*}
&&G_t+\w(i\d_x)G=0,\qquad\qquad -\infty<x<\infty,\quad t>0,\\
&&G(x,0)=\delta(x-x'),\qquad\quad\,\,\, -\infty<x<\infty,\quad0<x',\\
&&G(0,t)=0,\qquad\qquad\qquad\>\>\>\>\,\, t>0,
\end{eqnarray*}
and let $Q(x,t)$ satisfy
\begin{eqnarray*}
&&Q_t+\w(i\d_x)Q=0,\qquad\qquad -\infty<x<\infty,\quad t>0,\\
&&Q(x,0)=0,\qquad\quad\qquad\quad\,\,\,\,\, -\infty<x<\infty,\quad0<x',\\
&&Q(0,t)=g_0(t),\qquad\qquad\quad\>\>\,\, t>0.
\end{eqnarray*}
Note that both $G(x,x',t)$ and $Q(x,t)$ can be readily obtained by
means of the Fourier $x$-transform. Then, the solution to
Eqs. (\ref{d1}) is given by
\begin{equation}
\label{dsol}
q(x,t)=\int_0^{\infty}dx'\,\{G(x,x',t)-G(x,-x',t)\}q_0(x')+Q(x,t),\qquad x>0.
\end{equation}
Indeed, $G(x,x',t)$, $G(x,-x',t)$, $Q(x,t)$ satisfy Eq.~(\ref{d1}a)
and so does their linear superposition, Eq.~(\ref{dsol}). The initial
condition~(\ref{d1}b) is satisfied by the term involving $G(x,x',t)$,
since $G(x,-x',t)$ and $Q(x,t)$ do not contribute to $q(x,t)$ at
$t=0$. Finally, the boundary condition~(\ref{d1}c) is satisfied by
$Q(x,t)$ alone, since the terms involving $G$ cancel out at $x=0$.  For the latter to be true, it is crucial that $\w(\cdot)$ be an even polynomial of its argument.  For example, Eqs.~(\ref{1.2}) {\em cannot} be treated by the method of images, because $\w(l)=i(l+l^3)$ contains only odd powers of $l$.  Neither does the method of images apply for Eqs.~(\ref{1.1}), unless $\a=0$, since $\w(l)=-i\a l+l^2$.

Sometimes it is possible to apply the method of images {\em after} using a
suitable transformation. For example, such a transformation for
Eq.~(\ref{1.1}a) is
\begin{equation}
\label{1.7}
u(x,t)=q(x,t)e^{\a x/2}.
\end{equation}
Using this transformation, Eqs.~(\ref{1.1}) become
\begin{subequations}
\label{1.8}
\begin{eqnarray}
&&u_t=u_{xx}-\frac{\a^2}{4}u,\qquad\qquad 0<x<\infty,\quad t>0,\\
&&u(x,0)=q_0(x)e^{\a x/2},\qquad 0<x<\infty,\\
&&u(0,t)=g_0(t),\qquad\qquad\>\>\,\, t>0.
\end{eqnarray}
\end{subequations}
Eqs.~(\ref{1.8}) {\em can} be solved both by an $x$-sine transform and by
the method of images, provided that $u(x,0)$ decays as $x\to\infty$; since
the r.h.s. of Eq.~(\ref{1.8}b) involves $\exp(\a x/2)$ it follows that,
for a general initial condition $q_0(x)$, this is the case iff $\a<0$.

The method of images may also work for von Neumann boundary conditions. 
However, for mixed boundary conditions, such as $q_x(x,0)+\beta q(x,0)=g_0(t)$, the application of the method of images is far from straightforward.

\bigskip\noindent
{\bf Multidimensional IBV problems}
\medskip

It was noted earlier that IBV problems in $\{0<x<\infty,t>0\}$ {\em
cannot} in general be solved by an $x$-transform.  This is the case not
only for Eqs.~(\ref{1.2}), but apparently also for Eqs.~(\ref{1.1}) with
$\a>0$.  In this case, for problems in {\em one} spatial dimension one may
attempt to use the Laplace transform in $t$. However, even this approach
fails for {\em multidimensional} problems, since in this case one obtains
a PDE in the spatial dimensions, for which there {\em does not exist} a
proper transform.

\bigskip\noindent
{\bf The new method}
\medskip

For equations in one spatial dimension the new method constructs $q(x,t)$
as an integral in the complex $k$-plane, involving an $x$-transform of the
initial condition and a $t$-transform of the boundary conditions. For
equations in $N$ spatial dimensions the situation is similar, where the
integral representation is now constructed in the complex
$(k_1,\dots,k_N)$-planes.

\begin{figure}
\centerline{\psfig{file=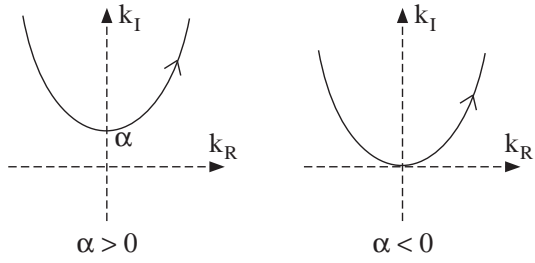,width=7cm,angle=0}}
\vspace{0.5cm}
\caption{The  contour $\d D_+$ associated with Eq.~(\ref{1.1}a).}
\label{fig1.1}
\end{figure}

For example, for the IBV problem~(\ref{1.1}), it will be shown in
section~II that this integral representation is
\begin{equation}
\label{1.9}
q(x,t)=\frac{1}{2\pi}\int_{-\infty}^{\infty} dk\,\e\qh_0(k)
+\frac{1}{2\pi}\int_{\d D_+}dk\,\e\gh(k,t),
\end{equation}
where $\w(k)=k^2-i\a k$, the oriented contour $\d D_+$ is the curve in the
complex $k$-plane defined by
\begin{equation}
\label{1.10}
k_I=\frac{\a}{2}+\sqrt{k_R^2+(\frac{\a}{2})^2},\qquad k=k_R+ik_I,
\end{equation}
see Fig.~\ref{fig1.1}, and the functions $\qh_0(k)$, $\gh(k,t)$ are defined in
terms of the given initial and boundary conditions as follows:
\begin{eqnarray}
\label{1.11}
&&\qh_0(k)=\int_0^{\infty}dx\,e^{-ikx}q_0(x),\qquad \Im k\leq0,\\
\label{1.12}
&&\gh(k,t)=-\qh_0(i\a-k)-(2ik+\a)\int_0^td\t\,e^{\w(k)\t}g_0(\t).
\end{eqnarray}
We note that $\gh(k,t)$ involves $\qh_0$ evaluated at $i\a-k$.

\bigskip\noindent
{\bf The long-time asymptotics}
\medskip

The representation obtained by the new method is convenient for computing
the long-time asymptotics of the solution:  Suppose that a given evolution
PDE is valid for $0<t<T$, where $T$ is a positive constant.  It can be
shown (see section~II) that the representation for $q(x,t)$ is equivalent
to the representation obtained by replacing $\gh(k,t)$ with $\gh(k,T)$. 
In particular, if $0<t<\infty$, then $\gh(k,t)$ can be replaced by
$\gh(k)\equiv\gh(k,\infty)$.  For example, the solution of the
IBV~(\ref{1.1}) with $0<t<\infty$ is given by~(\ref{1.9}) with $\gh(k,t)$
replaced by $\gh(k)$,
\begin{equation}
\label{1.13}
\gh(k)=-\qh_0(i\a-k)-(2ik+\a)\gh_0(k),\qquad
\gh_0(k)=\int_0^{\infty}dt\,e^{\w(k)t}g_0(t).
\end{equation}
Thus, the only time-dependence of $q(x,t)$ appears in the form
$\exp[ikx-\w(k)t]$; hence it is straightforward to obtain the long-time
asymptotics, using the steepest descent method. Similar considerations are
valid for evolution PDEs in multidimensions.

\bigskip\noindent
{\bf From the complex plane to the real axis}
\medskip

In case that the given IBV problem {\em can} be solved by an
$x$-transform, the relevant representation can be obtained by deforming
the integral representation of the solution obtained by the new method,
from the complex $k$-plane to the real axis.  Consider for example the
integral in the r.h.s. of Eq.~(\ref{1.9}) for the case that $\a<0$: It can
be verified that the functions $\exp(ikx)$, $\qh_0(i\a-k)$, and
$\exp[\w(k)(\t-t)]$ are bounded in the region of the complex $k$-plane
above the real axis and below the curve $\d D_+$.  Thus, the integral along
$\d D_+$ can be deformed to an integral along the real axis,
\begin{equation}
\label{1.14}
q(x,t)=
\frac{1}{2\pi}\int_{-\infty}^{\infty}dk\,\e\left[\qh_0(k)-\qh_0(i\a-k)
-(2ik+\a)\int_0^td\t\,e^{\w(k)\t}g_0(\t)\right].
\end{equation}
It can be shown that this formula is equivalent to the solution of
Eqs.~(\ref{1.7}), (\ref{1.8}) using the sine transform.  However, if
$\a>0$ this deformation is {\em not} possible.  Indeed, $\qh_0(i\a-k)$
involves $\exp[-ix(i\a-k)]=\exp[ik_Rx]\exp[-x(k_I-\a)]$; this term is
bounded for $k_I\geq\a$ (and in particular is bounded on $\d D_+$), but it
is {\em not} bounded in the region of the complex $k$-plane below the curve
$\d D_+$.

We note that even in the cases when it is possible to deform the contour
to the real axis, the representation in the complex $k$-plane has certain
advantages.  For example, the integral involving the sine transform is
{\em not} uniformly convergent at $x=0$ (unless $q(0,t)=0$).  Also, a
convenient way to study the long-time behavior of the representation
involving the sine transform is to {\em transform} it to the associated
representation in the complex $k$-plane.

\section{The Diffusion-Convection Equation}

\bigskip\noindent
{\bf Physical significance}
\medskip

Eq.~(\ref{1.1}a) arises, for example, from the diffusion-convection equation
\begin{equation}
\label{diff-conv}
q_{\tau}=Dq_{\xi\xi}+\a vq_{\xi},\quad \a=\pm1,\qquad0<\xi<\infty,\quad \tau>0,
\end{equation}
where $q(\xi,\tau)$ is a probability density function in the spatial and time variables $\xi$ and $\tau$,  $D$ is a diffusion coefficient with dimensions of $({\rm length})^2/({\rm time})$, and $v$ is a convection field  with dimensions of $({\rm length})/({\rm time})$.  Eq.~(\ref{1.1}a) is the normalized form of~(\ref{diff-conv}), expressed in terms of the dimensionless variables $t=(D/v^2)\tau$ and $x=(D/v)\xi$.  The dependence upon the  single spatial variable $x$ is justified in systems that are homogeneous in the transverse directions to $x$.  The choice $\a=+1$ corresponds to convection (a background drift velocity $v$) directed toward the origin, while $\a=-1$ corresponds to a background drift away from the origin.  In the long-time asymptotic limit convection dominates diffusion and determines the 
fate of the system. Suppose for example that the boundary condition is
$q(0,t)=0$, corresponding to an ideal sink at the origin, for the
case, of say, of particle flow.  Then, if $\a=+1$ the particles will flow to the sink at typical speed $v$, and the probability density is expected to decay to zero exponentially with time.  If $\a=-1$, the particles are drifting away from the sink; in this case one expects a depleted zone near the origin, which grows linearly with time, and exponential convergence to a finite level of survival.  The case of generic boundary conditions is harder to predict by such heuristic arguments.

\bigskip\noindent
{\bf The method of images}
\medskip

We have seen that the method of images is applicable in the singular case of $\alpha=0$, 
and even then the solution is straightforward only with pure Dirichlet
or von Neumann boundary conditions.  For $\a<0$ the method of images
can be used after a suitable transform, see Eqs.~(\ref{1.8}), and also
provided that the boundary conditions are simple enough.  For $\a>0$,
the method of images fails to provide a solution for generic initial
conditions.  We now show how the new method serves as a natural
extension of the methods of images and of classical integral transforms, even as these methods fail.  For pedagogical reasons, we expose the new method simultaneously for problems~(\ref{1.1}) and (\ref{1.2}).  The physical interpretation of~(\ref{1.2}) is discussed in Section~\ref{deVries}.

\bigskip\noindent
{\bf The new method}
\medskip

Suppose that a linear evolution PDE in one space variable admits the
solution $\exp[ikx-\w(k)t]$. For well posedness we assume that
\begin{equation}
\label{2.1}
\Re\w(k)>0, \qquad{\rm for\ } k {\rm\ real}.
\end{equation}  
The starting point of the new
method is to rewrite this PDE in the form
\begin{equation}
\label{2.2} 
\left(e^{-ikx+\omega(k)t} q\right)_t
+\left(e^{-ikx+\omega(k)t} X\right)_x=0,
\end{equation}
where
\begin{equation}
\label{2.3}
X=-i\left(\frac{\w(k)-\w(l)}{k-l}\right)q,\qquad l=-i\d_x.
\end{equation}
For Eq.~(\ref{1.1}a), $\w(k)=k^2-i\a k$, thus
\[
X=-i\left(\frac{k^2-l^2-i\a(k-l)}{k-l}\right)q=-[i(k+l)+\a]q.
\]
For Eq.~(\ref{1.2}a), $\w(k)=i(k+k^3)$, thus
\[
X=\left(\frac{(k-l)+(k^3-l^3)}{k-l}\right)q=(1+k^2+l^2+kl)q.
\]
Hence Eqs.~(\ref{1.1}a) and (\ref{1.2}a) can be written in the form
(\ref{2.2}), where $X$ is given, respectively, by
\begin{equation}
\label{2.4}
X=-q_x-(ik+\a)q,\qquad X=-q_{xx}-ikq_x+(1+k^2)q.
\end{equation}

\bigskip\noindent
(a) {\it The Fourier transform representation}
\medskip

The solution of Eq.~(\ref{2.2}) can be expressed in the form
\begin{equation}
\label{2.5}
q(x,t)=\frac{1}{2\pi}\int_{-\infty}^{\infty}dk\,\e\qh_0(k)
+\frac{1}{2\pi}\int_{-\infty}^{\infty}dk\,\e\gh(k,t),
\end{equation}
where
\begin{eqnarray}
\label{2.6}
&&\qh_0(k)=\int_0^{\infty}dx\,e^{-ikx}q_0(x),\qquad\Im k\leq0,\\
\label{2.7}
&&\gh(k,t)=\int_0^td\t\,e^{\w(k)\t}X(0,\t,k),\qquad k\in{\bf C}.
\end{eqnarray}
Furthermore, the functions $\qh_0(k)$ and $\gh(k,t)$ satisfy the {\em
global relation}
\begin{equation}
\label{2.8}
\gh(k,t)+\qh_0(k)=e^{\w(k)t}\qh(k,t),\qquad\Im k\leq0,
\end{equation}
where $\qh(k,t)$ denotes the $x$-Fourier transform of $q(x,t)$.

Indeed, using~(\ref{2.2}) it is straightforward to compute the time
evolution of $\qh(k,t)$,
\begin{eqnarray*}
\left(e^{\w(k)t}\qh(k,t)\right)_t&=&
\int_0^{\infty}dx\,\left(e^{-ikx+\w(k)t}q(x,t)\right)_t\\
&=&-\int_0^{\infty}dx\,\left(e^{-ikx+\w(k)t}X\right)_x
=e^{\w(k)t}X(0,t,k).\nonumber  
\end{eqnarray*}
Integrating this equation we find Eq.~(\ref{2.8}).  Solving
Eq.~(\ref{2.8}) for $\qh(k,t)$ and then using the inverse Fourier
transform, we find Eq.~(\ref{2.5}).

\bigskip\noindent
(b) {\it An integral representation in the complex {\rm k}-plane}
\medskip

\begin{figure}
\centerline{\psfig{file=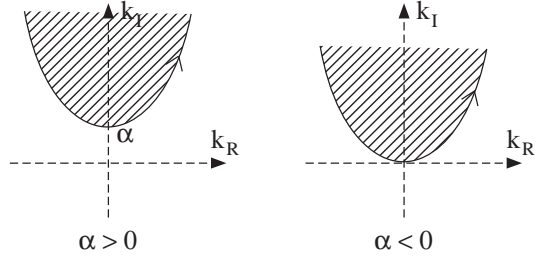,width=7cm,angle=0}}
\vspace{0.5cm}
\caption{The  domain $D_+$ for Eq.~(\ref{1.1}a).}
\label{fig2.1}
\end{figure}

\begin{figure}
\centerline{\psfig{file=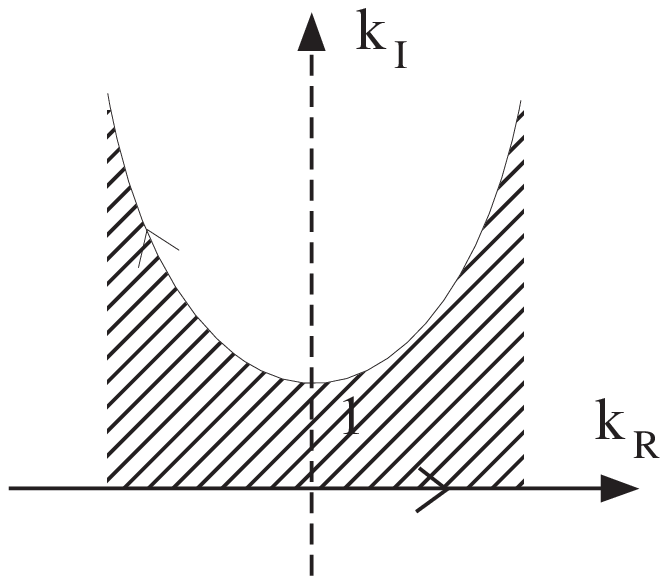,width=4cm,angle=0}}
\vspace{0.5cm}
\caption{The  domain $D_+$ for Eq.~(\ref{1.2}a).}
\label{fig2.2}
\end{figure}

The first crucial step of the new method is to replace the second integral
on the r.h.s. of Eq.~(\ref{2.5}) by an integral along the oriented curve
$\d D_+$.  This curve is the boundary of the domain $D_+$,
\begin{equation}
\label{2.9}
D_+=\{k\in{\bf C},\>\Im k>0,\>\Re\w(k)<0\},
\end{equation}
oriented so that $D_+$ is on the left of $\d D_+$.  For example, for
Eqs.~(\ref{1.1}a) and (\ref{1.2}a) the domains $D_+$ are the shaded regions
in Figs.~\ref{fig2.1} and \ref{fig2.2}, respectively.
Each of the curves in Fig.~\ref{fig2.1} is defined by Eq.~(\ref{1.10}), while the
curve in Fig.~\ref{fig2.2} is defined by
\[
k_I=\sqrt{1+3k_R^2}.
\]
Indeed, for Eq.~(\ref{1.1}a),
\[
\Re\w(k)=\Re\{(k_R+ik_I)^2-i\a(k_R+ik_I)\}=k_R^2-k_I^2+\a k_I.
\]
Thus, $D_+$ is the domain of the upper half complex $k$-plane specified by
\[
(k_I-\frac{\a}{2})^2-(k_R^2+(\frac{\a}{2})^2)>0.
\]
For Eq.~(\ref{1.2}a),
\[
\Re\w(k)=\Re\{i(k_R+ik_I)^3+i(k_R+ik_I)\}=k_I(k_I^2-3k_R^2-1).
\]
Thus, $D_+$ is the domain of the upper half complex $k$-plane specified by
\[
k_I^2-3k_R^2-1<0.
\]

The deformation of the integral from the real axis to the integral along
the curve $\d D_+$ is a direct consequence of Cauchy's theorem.  Indeed,
the term
\[
\e\gh(k,t)=e^{ikx}\int_0^td\t\,e^{-\w(k)(t-\t)}X(0,\t,k)
\]
is analytic and bounded in the domain $E_+$,
\[
E_+=\{k\in{\bf C},\>\Im k>0,\>\Re\w(k)>0\}.
\]
Thus, using Jordan's lemma in $E_+$, the integral along the boundary of
$E_+$ vanishes.

For example, for Eq.~(\ref{1.1}a) $E_+$ is the domain above the real axis
and below the curve $\d D_+$; thus the integral along the real axis can be
deformed to the integral along $\d D_+$.  For Eq.~(\ref{1.2}a), $E_+$ is
the domain above the curve $\{k_I=\sqrt{1+3k_R^2}\}$, thus the integral
along this curve vanishes; hence $\d D_+$ is the {\em union} of the
real axis and the curve $\{k_I=\sqrt{1+3k_R^2}\}$.

\bigskip\noindent
(c) {\it Analysis of the global relation}
\medskip

The second crucial step in the new method is the analysis of the global
relation (\ref{2.8}): This yields $\gh(k,t)$ in terms of $\qh_0(k)$ and 
the $t$-transform of the given boundary conditions.

Before implementing this step, we note that we have already used
Eq.~(\ref{2.8}) in the derivation of Eq.~(\ref{2.5}).  However, while
Eq.~(\ref{2.8}) was used earlier only for real $k$, in what follows it
will be used in the complex $k$-plane.

\begin{figure}
\centerline{\psfig{file=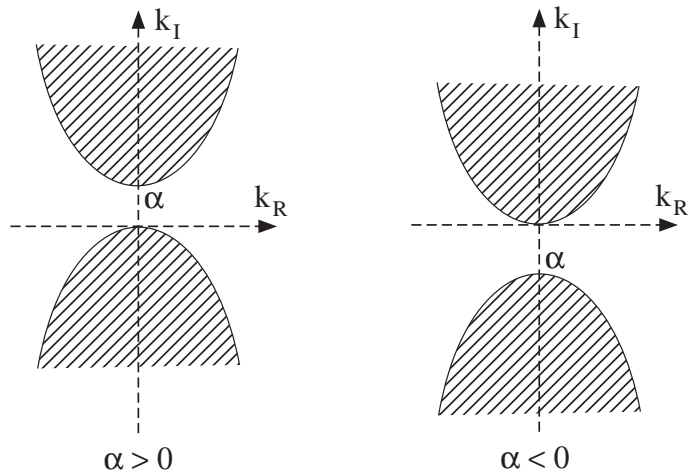,width=9cm,angle=0}}
\vspace{0.5cm}
\caption{The  domains $D_+$ and $D_-$ associated with Eq.~(\ref{1.1}a).}
\label{fig2.3}
\end{figure}

Recall that $\gh(k,t)$ is defined by Eq.~(\ref{2.7}), where for
Eq.~(\ref{1.1}a) $X$ is given by (\ref{2.4}a).  Thus
\begin {equation}
\label{2.10}
\gh(k,t)=-\gh_1(\w(k),t)-(ik+\a)\gh_0(\w(k),t),\qquad \w(k)=k^2-i\a k,
\end{equation}
where
\[
\gh_1(\w(k),t)=\int_0^td\t\,e^{\w(k)\t}q_x(0,\t),\quad
\gh_0(\w(k),t)=\int_0^td\t\,e^{\w(k)\t}g_0(\t);
\]
we have used the notation $\gh_1(\w(k),t)$ and $\gh_0(\w(k),t)$ to
emphasize that {\em  $\gh_1$ and $\gh_0$ depend on $k$ only through
$\w(k)$}.  Substituting the expression for $\gh(k,t)$ from~(\ref{2.10}) into the global relation (\ref{2.8}) we find
\begin{equation}
\label{2.11}
-\gh_1(\w(k),t)=(ik+\a)\gh_0(\w(k),t)-\qh_0(k)+e^{\w(k)t}\qh(k,t),\qquad\Im
k\leq0.
\end{equation}
Our task is to compute $\gh(k,t)$ on the curve $\d D_+$; since $\gh(k,t)$
is analytic in $D_+$ this is equivalent to computing $\gh(k,t)$ for $k\in
D_+$. Eq.~(\ref{2.10}) shows that $\gh(k,t)$ involves $\gh_0$, which is
{\em known} in terms of the given boundary condition $g_0(t)$, as well as
$\gh_1$, which involves the {\em unknown} boundary value $q_x(0,t)$.  In
order to compute $\gh_1$ using the global relation (\ref{2.11}), we first
transform Eq.~(\ref{2.11}) from the lower half complex $k$-plane to the
domain $D_+$.  In this respect, it is important to observe that since
$\gh_1$ depends on $k$ only through $\w(k)$, {\em $\gh_1$ remains
invariant by those transformations $k\to\nu(k)$ which preserve $\w(k)$}. 
For Eq.~(\ref{1.1}a), the equation $\w(k)=\w(\nu(k))$ has one non-trivial
root (the trivial root is $\nu(k)=k$),
\[
\nu^2-i\a\nu=k^2-i\a k,\qquad \nu(k)=i\a-k.
\]
Let $D_-$ be defined by
\begin{equation}
\label{2.D-}
D_-=\{k\in{\bf C},\> \Im k<0,\> \Re \w(k)<0\}.
\end{equation}
For $\w(k)=k^2-i\a k$, the domains $D_+$ and $D_-$ are the shaded regions
depicted in Fig.~\ref{fig2.3}.  The transformations that leave $\w(k)$ invariant
map the domain $\{D_+\oplus D_-\}$ onto itself.  Thus, if
$k\in D_+$, then $i\a-k\in D_-$.  Hence, replacing $k$ by $i\a-k$ in
Eq.~(\ref{2.11}), we find an equation that is valid for $k$ in $D_+$:
\begin{equation}
\label{2.12}
-\gh_1=-ik\gh_0-\qh_0(i\a-k)+e^{\w(k)t}\qh(i\a-k,t),\qquad k\in D_+.
\end{equation}
Replacing $-\gh_1$ in (\ref{2.10}) by the r.h.s. of the above equation, we
find
\[
\gh(k,t)=-(2ik+\a)\gh_0(\w(k),t)-\qh_0(i\a-k)+e^{\w(k)t}\qh(i\a-k,t).
\]
The term $e^{\omega(k)t} \qh(i\a-k,t)$ does {\em not} contribute to
$q(x,t)$. Indeed, this term gives rise to the integral
\[
\int_{\d D_+}dk\,e^{ikx}\qh(i\a-k,t),
\]
for which we note: $\exp(ikx)$ is bounded and analytic for $\Im k>0$; the
term $\qh(i\a-k,t)$ involves
$\exp[-i(i\a-k)x]=\exp(ik_Rx)\exp[-x(k_I-\a)]$, which is bounded and
analytic for $k_I>\a$, i.e., in $D_+$. Thus using Jordan's lemma in
$D_+$ it follows that the above integral vanishes, hence the effective
part of $\gh(k,t)$ is given by~(\ref{1.12}).

\bigskip\noindent
(d) {\it The long-time asymptotics}
\medskip

Suppose that a given evolution PDE is valid for $0<t<T$, where $T$ is a
positive constant.  Then, as mentioned earlier, we can replace
$\gh(k,t)$ by $\gh(k,T)$.  Indeed, the two representations associated with
$\gh(k,t)$ and $\gh(k,T)$ differ by
\[
\int_{\d D_+}dk\,\e\int_t^Td\t\,e^{\w(k)\t}X(0,\t,k).
\]
Since $T\geq t$, the coefficient $(\t-t)$ of $\w(k)$ is non-negative, thus
$\exp[ikx+\w(k)(\t-t)]$ is bounded and analytic in $D_+$, and Jordan's
lemma implies that the above integral vanishes.

In the case of Eq.~(\ref{1.1}a) the solution is given by
Eqs.~(\ref{1.9})-(\ref{1.11}), Eq.~(\ref{1.13}). The long-time
asymptotics  of $q(x,t)$ can be analyzed by the method of
steepest descent.  For $x=vt$, the critical point occurs at $d\w/dk=iv$, or
\[
k=\frac{i}{2}(\a+v)\equiv i\g .
\]
Let $v>\a>0$, then both integrals on the r.h.s. of (\ref{1.9})
contribute  to the long-time asymptotics, since the path of
integration can be modified to pass through the critical point
in each case. The behaviour of the argument of the relevant
exponentials near the critical point is given by
\[
ikvt-\w(k)t=-\g^2t-(k-i\g)^2t,
\]
hence the leading-order asymptotics to $q(vt,t)$ is
\[
\frac{1}{2\pi}\qh_0(i\g)e^{-\g^2t}\int_{-\infty}^{\infty}
  e^{-(k-i\g)^2t}dk
+\frac{1}{2\pi}\gh(i\g)e^{-\g^2t}\int_{\d D_+}
  e^{-(k-i\g)^2t}dk
=\frac{1}{2\sqrt{\pi t}}[\qh_0(i\g)+\gh(i\g)]e^{-\g^2t}.
\]
The numerical value of $\qh_0(i\g)$ and $\gh(i\g)$ are given explicitly in terms
of the initial and boundary conditions, see Eqs.~(\ref{1.11}) and (\ref{1.13}).

For $\a>v>0$ the second integral on the r.h.s. of (\ref{1.9}) does not contribute,
since the critical point lies outside of the domain $D_+$.  In this case the
leading-order asymptotics is
\[
q(vt,t)\sim\frac{1}{2\sqrt{\pi t}}\qh_0(i\g)e^{-\g^2t}.
\]

For $\a<0$ and $v>-\a$ both integrals contribute and the answer is
identical with that of the
case $v>\a>0$.  Finally, for $\a<0$ and $v<-\a$ neither integral contributes and the 
leading-order asymptotic behavior is zero.  The latter is expected on physical grounds, 
since the (rightward) drift sweeps the probability density $q(vt,t)$ past $x=vt$. 

\section{The Linearized Korteweg - de\ Vries Equation}
\label{deVries}

Eq.~(\ref{1.2}a) is the linear limit of the celebrated Korteweg - de\ Vries
equation:
\begin{equation}
\label{3.1}
q_t+q_x+\lambda q_{xxx}+6qq_x=0,\quad \lambda=\pm1,
\end{equation}
for the case of $\lambda=-1$; this corresponds to dominant surface
tension.  Eq.~(\ref{3.1}) is the normalized form of 
\begin{equation}
\label{3.2}
\frac{\d\eta}{\d\tau}=\frac{3}{2}\sqrt{\frac{g}{h}}\frac{\d}{\d\xi}
(\frac{1}{2}\eta^2+\eta+\frac{1}{3}\sigma\frac{\d^2\eta}{\d\xi^2}),
\qquad \sigma=\frac{1}{8}h^3-\frac{Th}{\rho g},
\end{equation}
where $\eta$ is the elevation of the water above the equilibrium level $h$,
$T$ is the surface tension, $\rho$ is the density of the medium, and $g$
is the free-fall acceleration constant.  
This equation is the small amplitude, long wave
limit of the equations describing idealized (inviscid) water waves under
the assumption of irrotationality.
Eq.~(\ref{3.1}) is obtained
after transforming to the dimensionless variables 
$t=\frac{1}{2}\sqrt{\frac{g}{h\sigma}}\tau$, $x=-\sigma^{-1/2}\xi$, 
$q=\eta/2$.

Eq.~(\ref{3.1}) usually appears without the $q_x$ term.  This is because
the Korteweg - de\ Vries equation is usually studied on the full line and then
the term $q_x$ can be eliminated by means of a Galilean
transformation.  However, for the half-line this transformation would
change the domain from a quarter-plane to a wedge.

Laboratory experiments with water waves typically involve Eq.~(\ref{3.1})
with $\lambda=1$, $q(x,0)=0$, and $q(0,t)$ a periodic function of $t$. 
Thus the linearized version of Eq.~(\ref{3.1}) with $\lambda=1$ is valid
until {\em small} amplitude waves reach the opposite end of the water
tank.  Eq.~(\ref{1.2}a) is valid under similar circumstances, in the case
of dominant surface tension.

It is interesting to note that while in the case of $\lambda=1$ the
problem is well posed with only {\em one} boundary condition at $x=0$, in
the case of $\lambda=-1$ the problem is well posed with {\em two} boundary
conditions. The case of $\lambda=1$ is solved in~\cite{IMA}.  Here we
solve the case of $\lambda=-1$.

\begin{figure}
\centerline{\psfig{file=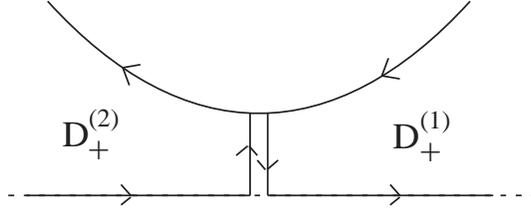,width=7cm,angle=0}}
\vspace{0.5cm}
\caption{The  contours $\d D_+^{(1)}$ and $\d D_+^{(2)}$ associated with Eq.~(\ref{1.2}a).}
\label{fig3.1}
\end{figure}

The solution of the IBV problem~(\ref{1.2}) is given by Eq.~(\ref{1.9}),
where $\w(k)=i(k+k^3)$, $\qh_0(k)$ is the Fourier transform of $q_0(x)$
(see Eq.~(\ref{1.11})), $\gh(k,t)$ is defined by
\begin{eqnarray}
\label{3.3}
&&\gh(k,t)=-\qh_0(\nu)+i(\nu-k)\int_0^td\t\,e^{\w(k)\t}g_1(\t)
 +(k^2-\nu^2)\int_0^td\t\,e^{\w(k)\t}g_0(\t),\\
&&\nu=\frac{-k-i\sqrt{3k^2+4}}{2}\>\>{\rm for\ }k\in D_+^{(1)},\qquad
\nu=\frac{-k+i\sqrt{3k^2+4}}{2}\>\>{\rm for\ }k\in D_+^{(2)},\nonumber
\end{eqnarray}
and $\d D_+$ is the union of the boundaries of $D_+^{(1)}$ and $D_+^{(2)}$
depicted in Fig.~\ref{fig3.1} (the relevant curve is
$\{k_I=\sqrt{3k_R^2+1},\,k_I>0\}$).

\begin{figure}
\centerline{\psfig{file=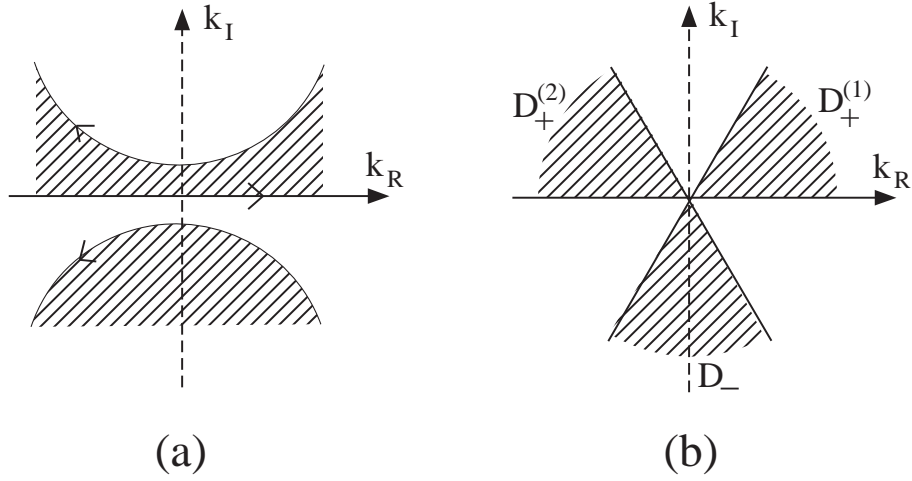,width=12cm,angle=0}}
\vspace{0.5cm}
\caption{The  domains $D_+$ and $D_-$ for Eq.~(\ref{1.2}a).}
\label{fig3.2}
\end{figure}

Indeed, the representation (\ref{1.9}) was derived in section~II.  Using
the definition of $\gh(k,t)$, Eq.~(\ref{2.7}), and recalling that $X$ is
given by Eq.~(\ref{2.4}b), we find
\begin{equation}
\label{3.4}
\gh(k,t)=-\gh_2(\w(k),t)-ik\gh_1(\w(k),t)+(1+k^2)\gh_0(\w(k),t),\qquad
\w(k)=i(k+k^3),
\end{equation}
where $\gh_1$, $\gh_0$ are the first, second integrals appearing in the
r.h.s. of Eq.~(\ref{3.3}) and $\gh_2$ involves the unknown boundary value
$q_{xx}(0,t)$,
\[
\gh_2(\w(k),t)=\int_0^td\t\,e^{\w(k)\t}q_{xx}(0,\t).
\]
Substituting the expression for $\gh(k,t)$ from Eq.~(\ref{3.4}) into the
global relation~(\ref{2.8}), we find
\begin{equation}
\label{3.5}
-\gh_2(\w(k),t)=ik\gh_1(\w(k),t)-(1+k^2)\gh_0(\w(k),t)-\qh_0(k)+e^{\w(k)t}\qh(k,t),
\qquad\Im k\leq0.
\end{equation}
The equation $\w(k)=\w(\nu(k))$ has two nontrivial roots,
\begin{eqnarray}
\label{3.6}
&&\nu+\nu^3=k+k^3,\qquad\nu^2+k\nu+k^2+1=0,
\qquad\nu_{1,2}=\frac{-k\mp i\sqrt{3k^2+4}}{2};\\
&&\nu_1(k)\sim e^{4i\pi/3}k,\qquad\nu_2(k)\sim e^{2i\pi/3}k,
\qquad k\to\infty.\nonumber
\end{eqnarray}
The domains $D_+$ and $D_-$ (see Eqs.~(\ref{2.9}) and (\ref{2.D-})) are
the shaded regions depicted in Fig.~\ref{fig3.2}a, where the two relevant curves
are $k_I=\pm\sqrt{3k_R^2+1}$;
the limit of the domains as
$k\to\infty$ are depicted in Fig.~\ref{fig3.2}b, where each of the relevant angles
is $\pi/3$.

If $\frac{2\pi}{3}<\arg k<\frac{4\pi}{3}$ then $
\frac{4\pi}{3}<\arg k+\frac{2\pi}{3}<2\pi$, and
$2\pi<\arg k+\frac{4\pi}{3}<\frac{8\pi}{3}$. Thus
\[
k\in D_+^{(2)},\qquad \nu_2(k)\in D_-,\qquad \nu_1(k)\in D_+^{(1)}.
\]
Similarly,
\[
k\in D_+^{(1)},\qquad \nu_2(k)\in D_+^{(2)},\qquad \nu_1(k)\in D_-.
\]
Thus, replacing $k$ by $\nu_1(k)$ and $\nu_2(k)$ in (\ref{3.5}), we find
\begin{eqnarray*}
&&-\gh_2=i\nu_1\gh_1-(1+\nu_1^2)\gh_0-\qh_0(\nu_1)+e^{\w(k)t}\qh(\nu_1(k),t),
\qquad k\in D_+^{(1)};\\
&&-\gh_2=i\nu_2\gh_1-(1+\nu_2^2)\gh_0-\qh_0(\nu_2)+e^{\w(k)t}\qh(\nu_2(k),t),
\qquad k\in D_+^{(2)}.
\end{eqnarray*}
Replacing $-\gh_2$ in (\ref{3.4}) by the r.h.s. of the above equations, we
find
\[
\gh(k,t)=i(\nu-k)\gh_1+(k^2-\nu^2)\gh_0-\qh_0(\nu)+e^{\w(k)t}\qh(\nu,t),
\qquad k\in D_+.
\]
Due to analiticity considerations the term $\qh(\nu,t)$ does not
contribute to $q(x,t)$, see section~II; thus the effective part of
$\gh(k,t)$ is given by~(\ref{3.3}).

\bigskip\noindent
{\bf The long-time asymptotics}
\medskip

The long-time asymptotics can again be analyzed by replacing $\gh(k,t)$ with
$\gh(k,\infty)=\gh(k)$ and following standard steepest descent or stationary phase
expansions.  One thus finds~\cite{F-Schultz}: For $v=x/t>1$, $q(vt,t)$ satisfies
\[
q(vt,t)=\frac{1}{\sqrt{12\pi\g t}}\big[(P(\g)-P(\nu(\g))e^{i\phi}
  +(P(-\g)-P(\nu(-\g))e^{-i\phi}\big]+\O(t^{-3/2}),
\]
as $t\to\infty$, where
\begin{eqnarray*}
\g&=&\sqrt{\frac{v-1}{3}},\\
\phi(t)&=&2\g^3t-\pi/4,\\
\nu(k)&=&-\frac{1}{2}(k+i\sqrt{3k^2+4}),\\
P(k)&=&\qh_0(k)+(k^2+1)\gh_0(k)-ik\gh_1(k),
\end{eqnarray*}
$\qh_0(k)$ is defined in terms of the initial condition (Eq.~(\ref{1.13})), and 
$\gh_1(k)$, $\gh_0(k)$ are the first, second integrals appearing in the r.h.s. of 
Eq.~(\ref{3.3}) evaluated with $t\to\infty$.  For $v<1$, $q(vt,t)$ decays faster than any algebraic power of $t$, as $t\to\infty$.

\section{The Diffusion Equation in Multidimensions}

\bigskip\noindent
{\bf Physical significance}
\medskip

This problem is the generalization of problem (\ref{1.1}) to multidimensions.
It arises from the diffusion-convection equation
\begin{equation}
\label{D3.1}
q_{\t}=\sum_{j=1}^N (Dq_{\xi_j\xi_j}+v_jq_{\xi_j}), \qquad 0<\xi_j<\infty,\quad\tau>0,\quad
j=1,2,\dots N,
\end{equation}
which represent isotropic diffusion in a quadrant of $N$-dimensional space, 
with diffusion
coefficient $D$, and under a background convection (drift) field $(v_1,v_2,\dots,v_N)$.
Eq.~(\ref{1.3}a) is obtained upon passing to the dimensionless variables 
$t=(D/{\bar v}^2)\t$ and $x_j=(D/{\bar v})\xi_j$, $j=1,2,\dots,N$, 
where ${\bar v}$ is a typical speed such 
as the rms ${\bar v}=(\sum_jv_j^2)^{1/2}$.

\bigskip\noindent
{\bf The new method}
\medskip

For the solution of the IBV problem~(\ref{1.3}) we introduce the following
notations:
\begin{eqnarray*}
&&\bullet\quad
x=(x_1,\dots,x_N),\qquad k=(k_1,\dots,k_N),\qquad 
k\cdot x=\sum_1^Nk_lx_l,\qquad 
k^{(j)}\cdot x^{(j)}=\sum_{l=1,l\neq j}^N k_lx_l.\\
&&\bullet\quad
\int_{{\bf
R}^N}dk=\int_{-\infty}^{\infty}dk_1\cdots\int_{-\infty}^{\infty}dk_N.\\
&&\bullet\quad
C_j \text{\ is the curve\ }
\{(k_j)_I>0,\>(k_j)_I=\frac{\a_j}{2}+\sqrt{[(k_j)_R]^2+(\frac{\a_j}{2})^2},
\>j=1,2,\dots,N\}.\\
&&\bullet\quad
\nu_j=i\a_j-k_j,\qquad j=1,2,\dots,N.\\
&&\bullet\quad
\w(k)=\sum_{j=1}^N(k_j^2-i\a_jk_j).\\
&&\bullet\quad
\qh_0(k)=\int_0^{\infty}dx_1\cdots\int_0^{\infty}dx_N\, e^{-ik\cdot
x}q_0(x).\\
&&\bullet\quad
\gh_0^{(j)}(k,t)=\int_0^td\t\int_0^{\infty}dx_1\cdots\int_0^{\infty}
dx_{j-1}
 \int_0^{\infty} dx_{j+1}\cdots\int_0^{\infty}dx_N\,e^{-ik^{(j)}\cdot
x^{(j)}+\w(k)\t}
  g_0^{(j)}(x_1,\dots,x_{j-1},x_{j+1},\dots,x_N,\t),\\
&&\quad\>\> j=1,2,\dots,N.\\
&&\bullet\quad
\Qh(k,t)=\qh_0(k)-\sum_1^N(ik_j+\a_j)\gh_0^{(j)}(k,t).
\end{eqnarray*}

The solution of the IBV problem (\ref{1.3}) is given by
\begin{eqnarray}
\label{4.1}
q(x,t)=&&\frac{1}{(2\pi)^N}\int_{{\bf R}^N}dk\,\edot\qh_0(k)\nonumber\\
&&-\frac{1}{(2\pi)^N}\sum_1^N
 \int_{-\infty}^{\infty}dk_1\cdots\int_{-\infty}^{\infty} dk_{j-1}
 \int_{C_j} dk_j
 \int_{-\infty}^{\infty} dk_{j+1}\cdots\int_{-\infty}^{\infty}dk_N\,\edot
  (ik_j+\a_j)\gh_0^{(j)}(k,t) \nonumber\\
&&-\frac{1}{(2\pi)^N}\left(
 \int_{C_1}dk_1\int_{-\infty}^{\infty}dk_2\cdots\int_{-\infty}^{\infty}dk_N
  \edot\Qh(\nu_1,k_2,\dots,k_N)+CP \right)\nonumber\\
&&+\frac{1}{(2\pi)^N}\left(
\int_{C_1}dk_1\int_{C_2}dk_2\int_{-\infty}^{\infty}dk_3\cdots\int_{-\infty}^{\infty}dk_N
  \edot\Qh(\nu_1,\nu_2,k_3,\dots,k_N)+CP \right)\nonumber\\
&&-\cdots+
\frac{(-1)^N}{(2\pi)^N}
 \int_{C_1}dk_1\int_{C_2}dk_2\cdots\int_{C_N}dk_N\,
  \edot\Qh(\nu_1,\nu_2,\dots,\nu_N),
\end{eqnarray}
where $CP$ denotes cyclic permutations (i.e., all possible relevant
combinations).

For the derivation of Eq.~(\ref{4.1}) we introduce the additional
notations:
\begin{eqnarray*}
&&\bullet\quad
x^{(j)}=(x_1,\dots,x_{j-1},x_{j+1},\dots,x_N),\qquad
k^{(j)}=(k_1,\dots,k_{j-1},k_{j+1},\dots,k_N).
\\ &&\bullet\quad
\int_{{\bf
R}_+^N}dx=\int_{0}^{\infty}dx_1\cdots\int_{0}^{\infty}dx_N.\\
\\ &&\bullet\quad
\int_{{\bf
R}_+^{(j)}}dx^{(j)}=\int_{0}^{\infty}dx_1\cdots
\int_{0}^{\infty}dx_{j-1}\int_{0}^{\infty}dx_{j+1}
\cdots\int_{0}^{\infty}dx_N.\\
\\ &&\bullet\quad
\int_{{\bf
R}^{(j)}}dk^{(j)}=\int_{-\infty}^{\infty}dk_1\cdots
\int_{-\infty}^{\infty}dk_{j-1}\int_{-\infty}^{\infty}dk_{j+1}
\cdots\int_{-\infty}^{\infty}dk_N.\\
&&\bullet\quad
e=\edot.\\
&&\bullet\quad
E=e^{\w(k)t}.
\end{eqnarray*}

Substituting $\exp[ik\cdot x-\w(k)t]$ in Eq.~(\ref{1.3}a) we find that
$\w(k)$ is given by the expression defined in the notations.

Eq.~(\ref{1.3}a) can be written in the form
\begin{equation}
\label{4.2}
\left(e^{-ik\cdot x+\w(k)t}q\right)_t+
\sum_1^N\left(e^{-ik\cdot x+\w(k)t}X_j\right)_{x_j}=0,
\end{equation}
where \[
X_j=-q_{x_j}-(ik_j+\a_j)q,\qquad j=1,2,\dots,N.
\]
Let $\qh(k,t)$ denote the $N$-dimensional Fourier transform of $q(x,t)$. 
Using Eq.~(\ref{4.2}) we find
\begin{eqnarray*}
\left(e^{\w(k)t}\qh(k,t)\right)_t&&=\int_{{\bf R}_+^N}dx\,\Big(
 e^{-ik\cdot x+\w(k)t}q(x,t)\Big)_t\\
&&=-\sum_{j=1}^N\int_0^{\infty}dx_j\Big(
 e^{-ik\cdot x+\w(k)t}X_j\Big)_{x_j}=
\sum_{j=1}^N e^{-ik\cdot x+\w(k)t}X_j|_{x_j=0}.
\end{eqnarray*}
Integrating this equation with respect to $t$, we obtain
\begin{equation}
\label{4.3}
\qh_0(k)+\sum_1^N\gh^{(j)}(k,t)=e^{\w(k)t}\qh(k,t),
\end{equation}
where $\qh_0(k)$ is the $N$-dimensional Fourier transform of $q_0(x)$ and
$\gh^{(j)}(k,t)$ is defined as follows:
\[
\gh^{(j)}(k,t)=-\int_0^td\t\int_{{\bf R}_+^{(j)}}dx^{(j)}\,
 e^{-ik^{(j)}\cdot x^{(j)}+\w(k)\t}[q_{x_j}(x^{(j)},\t)
  +(ik_j+\a_j)q(x^{(j)},\t)],
\]
or
\begin{equation}
\label{4.4}
\gh^{(j)}(k,t)\equiv\gh_1^{(j)}(-ik^{(j)},\w(k),t)-
(ik_j+\a_j)\gh_0^{(j)}(-ik^{(j)},\w(k),t),\qquad j=1,2,\dots,N,
\end{equation}
where $\gh_0^{(j)}$ is defined in the notations and $\gh_1^{(j)}$ denotes
the integral of $-q_{x_j}$.

Solving Eq.~(\ref{4.3}) for $\qh(k,t)$ and using the inverse
$N$-dimensional Fourier transform, we find
\begin{equation}
\label{4.5}
q(x,t)=\frac{1}{(2\pi)^N}\int_{{\bf R}^N}dk\,\edot\qh_0(k)
+\frac{1}{(2\pi)^N}\sum_1^N\int_{{\bf R}^N}dk\,\edot
  [\gh_1^{(j)}-(ik_j+\a_j)\gh_0^{(j)}].
\end{equation}
The terms $\gh_1^{(j)}$ and $\gh_0^{(j)}$ contain $k_j$ only through
$\w(k)$, thus the relevant integral in the complex $k_j$-plane can be
deformed from the real axis to the curve $C_j$ (see the discussion in
section~II).  Hence Eq.~(\ref{4.5}) yields the first two integrals
appearing in (\ref{4.1}) plus the term
\begin{equation}
\label{4.6}
\frac{1}{(2\pi)^N}\sum_1^N\int_{{\bf R}^{(j)}}dk^{(j)}\int_{C_j}dk_j\,\edot
 \gh_1^{(j)}.
\end{equation}
Using the global relation (\ref{4.3}), it can be shown that the above term
yields the remaining expressions appearing in (\ref{4.1}).  For
pedagogical reasons, we give the details for $N=3$:  In this case the
global condition becomes
\begin{equation}
\label{4.7}
\gh_1^{(1)}(-ik_2,-ik_3)+\gh_1^{(2)}(-ik_1,-ik_3)+\gh_1^{(3)}(-ik_2,-ik_3)=
-\Qh(k_1,k_2,k_3)+E\qh(k_1,k_2,k_3),\qquad \Im k_j\leq0,
\end{equation}
where $E$, $\Qh$ are defined in the notations and, for simplicity of
notation, we have dropped the $t$- and $\w(k)$-dependence.
Since $\gh_1^{(1)}$ in Eq.~(\ref{4.6}) is integrated along $C_1$, we
replace $k_1$ by $\nu_1$ in Eq.~(\ref{4.7}) and then solve the resulting
equation for $\gh_1^{(1)}(-ik_2,-ik_3)$.  Similarly, for $\gh_1^{(2)}$,
$\gh_1^{(3)}$, we replace $k_2$, $k_3$ in Eq.~(\ref{4.7}) by $\nu_2$,
$\nu_3$.  Thus the expression (\ref{4.6}) becomes 
\begin{eqnarray}
\label{4.8}
&&-\int_{{\bf R}^2}dk_2dk_3\int_{C_1}dk_1\,e[\gh_1^{(2)}(-i\nu_1,-ik_3)
 +\gh_1^{(3)}(-i\nu_1,-ik_2)]\nonumber\\
&&-\int_{{\bf R}^2}dk_1dk_3\int_{C_2}dk_2\,e[\gh_1^{(1)}(-i\nu_2,-ik_3)
 +\gh_1^{(3)}(-ik_1,-i\nu_2)]\nonumber\\
&&-\int_{{\bf R}^2}dk_1dk_2\int_{C_3}dk_3\,e[\gh_1^{(1)}(-ik_2,-i\nu_3)
 +\gh_1^{(2)}(-ik_1,-i\nu_3)]\\
&&-\int_{{\bf R}^2}dk_2dk_3\int_{C_1}dk_1\,e\Qh(\nu_1,k_2,k_3)
-\int_{{\bf R}^2}dk_1dk_3\int_{C_2}dk_2\,e\Qh(k_1,\nu_2,k_3)
-\int_{{\bf R}^2}dk_1dk_2\int_{C_3}dk_3\,e\Qh(k_1,k_2,\nu_3).\nonumber
\end{eqnarray}
In addition to these terms we obtain three terms involving
$\qh$, but these terms vanish due to analyticity considerations; for
example, one of these terms is
\[
\int_{{\bf R}^2}dk_2dk_3\int_{C_1}dk_1\,e^{ik\cdot
x}\qh(\nu_1,k_2,k_3,t),
\]
which vanishes since both $e^{ik\cdot x}$ and $\qh(\nu_1,k_2,k_3,t)$ are
bounded and analytic in the region of the complex $k_1$-plane above the
curve $C_1$.

The last 3 terms in ~(\ref{4.8}) are part of the expression for $q(x,t)$;
regarding the first 3 terms in~(\ref{4.8}), we note the following: the two
integrals involving $\gh^{(2)}$ contain $k_2$ only through $\w(k)$, thus
for these integrals the contour with respect to $k_2$ can be deformed from
the real $k_2$-axis to the curve $C_2$; similarly for the integrals
involving $\gh^{(1)}$ and $\gh^{(3)}$.  Thus we obtain
\begin{equation}
\label{4.9}
-\int_{{\bf R}}dk_3\int_{C_2}dk_2\int_{C_1}dk_1\,e
 [\gh_1^{(1)}(-i\nu_2,-ik_3)+\gh_1^{(2)}(-i\nu_1,-ik_3)],
\end{equation}
plus two more similar integrals, which can be obtained from~(\ref{4.9}) by
cyclic permutation $1\to2\to3\to1$.  The square bracket appearing
in~(\ref{4.9}) can be expressed in terms of $\gh_1^{(3)}$ using the global
relation evaluated at $k_1\to\nu_1$, $k_2\to\nu_2$,
\[
-\gh_1^{(1)}(-i\nu_2,-ik_3)-\gh_1^{(2)}(-i\nu_1,-ik_3)=
\gh_1^{(3)}(-i\nu_1,-i\nu_2)+\Qh(\nu_1,\nu_2,k_3)-E\qh(\nu_1,\nu_2,k_3).
\]
Due to analyticity considerations the term involving $\qh$ vanishes, thus
the expression~(\ref{4.9}) together with the other two similar expressions
yield
\begin{eqnarray}
\label{4.10}
&&\int_{\bf
R}dk_3\int_{C_2}dk_2\int_{C_1}dk_1\,e\gh_1^{(3)}(-i\nu_1,-i\nu_2)
+\int_{\bf
R}dk_2\int_{C_3}dk_3\int_{C_1}dk_1\,e\gh_1^{(2)}(-i\nu_1,-i\nu_3)
\nonumber\\
&&+\int_{\bf
R}dk_1\int_{C_2}dk_2\int_{C_3}dk_3\,e\gh_1^{(1)}(-i\nu_2,-i\nu_3)
\nonumber\\
&&+\int_{\bf
R}dk_3\int_{C_2}dk_2\int_{C_1}dk_1\,e\Qh(\nu_1,\nu_2,k_3)
+\int_{\bf
R}dk_2\int_{C_3}dk_3\int_{C_1}dk_1\,e\Qh(\nu_1,k_2,\nu_3)
\nonumber\\
&&+\int_{\bf
R}dk_1\int_{C_2}dk_2\int_{C_3}dk_3\,e\Qh(k_1,\nu_2,\nu_3).
\end{eqnarray}
The last three terms above, are part of the expression for $q(x,t)$;
regarding the first three terms we note the following: the integral
involving
$\gh^{(3)}$ contains $k_3$ only through $\w(k)$, thus the integration along
the real $k_3$-axis can be deformed to the curve $C_3$; similarly for the
other two integrals.  Thus the first three terms of~(\ref{4.10}) yield
\begin{equation}
\label{4.11}
\int_{C_1}dk_1\int_{C_2}dk_2\int_{C_3}dk_3\,e
[\gh_1^{(1)}(-i\nu_2,-i\nu_3)+\gh_1^{(2)}(-i\nu_1,-i\nu_3)
+\gh_1^{(3)}(-i\nu_1,-i\nu_2)].
\end{equation}
Using the global relation evaluated at $k_1\to\nu_1$, $k_2\to\nu_2$,
$k_3\to\nu_3$, the square bracket appearing in~(\ref{4.11}) can be
replaced by
\[
-\Qh(\nu_1,\nu_2,\nu_3)+E\qh(\nu_1,\nu_2,\nu_3,t).
\]
Thus the expression (\ref{4.11}) yields
\[
-\int_{C_1}dk_1\int_{C_2}dk_2\int_{C_3}dk_3\,e\Qh(\nu_1,\nu_2,\nu_3)
\]
plus a term involving $\qh$ that vanishes due to analiticity
considerations.

\bigskip\noindent
{\bf The long-time asymptotics}
\medskip

The long-time asymptotics may be analyzed by the method of steepest descent, after
replacing the variables $\gh^{(j)}(k,t)$ with $\gh^{(j)}(k)=\gh^{(j)}(k,\infty)$.

\acknowledgments
We gratefully acknowledge partial support of this work from the
NSF, under contract no.\ PHY-0140094 (D.b.-A.), and the EPRSC (A.S.F.).



\begin{thebibliography}{99}
\bibitem{remark} For simplicity, we are assuming that $\w(\cdot)$ is a polynomial of order less or equal 2.  When $\w(\cdot)$ is a polynomial of order $2n$ one needs $n$ boundary conditions and the method of images is complicated further.
\end{thebibliography}
\end{document}